\documentclass[twocolumn,showpacs,preprintnumbers,amsmath,amssymb]{revtex4}

\usepackage{graphicx}
\usepackage{dcolumn}
\usepackage{bm}

\begin{document}

\title{A Model of Weighted Network: the Student Relationships in a Class}
\author{Bo Hu}
\author{Xin-Yu Jiang}
\author{Jun-Feng Ding}
\author{Yan-Bo Xie}
\author{Bing-Hong Wang}
\email{bhwang@ustc.edu.cn, Fax:+86-551-3603574.}
\affiliation{%
Nonlinear Science Center and Department of Modern Physics,
University of Science and Technology of China, Hefei, 230026, PR
China \\
}%

\date{\today}

\begin{abstract}
A simple model is proposed to simulate the evolution of
interpersonal relationships in a class. The small social network
is simply assumed as an undirected and weighted graph, in which
students are represented by vertices, and the extent of favor or
disfavor between two of them are denoted by the weight of
corresponding edge. Various weight distributions have been found
by choosing different initial configurations. Analysis and
experimental results reveal that the effect of first impressions
has a crucial influence on the final weight distribution. The
system also exhibits a phase transition in the final hostility
(negative weights) proportion depending on the initial amity
(positive weights) proportion.

\end{abstract}


\maketitle

\section{Introduction}
Recently, physicists have displayed much interests in social
phenomena that exhibit complex behaviors and nonlinear dynamics. A
social network is a set of people with some pattern of contacts or
interactions between them \cite{ref1,ref2}. The patterns of
friendships between individuals \cite{ref3,ref4}, business
relationships between corporations \cite{ref5,ref6}, and
intermarriages between families \cite{ref7} are all examples of
networks that have been studied in the past. Hidden behind such
complex phenomena, however, are many factors hard to control,
including human nature, social environment, social distance and
opportunity. Fortunately, human beings have accumulated precious
experiences of themselves. Sociologists and psychologists have
long noticed that the first impression between two individuals is
often the seed of their future relationship, not only seen in
romantic stories. Good seed may promise good harvest, while ill
seed may be supposed to portend illness. This effect seems to be
more distinct in campus life, where the influence of social
distances is not so remarkable. Pupils of a class serve as a
typical example that exhibits relatively simple friendships. See
the studies of friendship networks of school children by Rapoport
\cite{ref4}. According to empirical observation, students are more
likely to get along with their friends in daily activities, such
as dinner, discussion, entertainment, etc. Often, their ties are
strengthened through frequent contacts. Assumably, people with
common friends or common ``enemies" are prone to unite; likewise,
the ``enemy" of Jack's friends or the friend of Jack's ``enemies"
may be very difficult to associate with Jack. Similar human
relations and social environments are an effective catalyzer for
friendships. As a result of restricted social scope and in the
light of psychology, the encounters between ``enemies" in a class
may also become quite frequent. You cannot avoid your foes in such
a small world. It might be an unfriendly eyesight, a provocative
action or an unexpected quarrel, as commonplace in daily life. In
the deepest of one's heart are always those he hates or he loves,
while people without intensified contact are easy to fade from
memory.

In this paper, a simple model is proposed to study the mechanism
of interpersonal relationships within a class. The small social
world is assumed as an undirected and weighted graph, where
students are represented by vertices, and the depth of favor or
disfavor between two are denoted by the weight of corresponding
edge. It must be stressed that this model is not restricted in
pupil relations, but can be applied to other cases such as
interpersonal relations in a club or a team. This paper is
organized as follows. In section 2, the model is described on the
basis of some simplified assumptions; to understand the mechanism
better, we analyze the case for $N$=3. Next, experimental results
of the weight distribution are presented and the first impression
effect is discussed. In section 4, we conclude with some outlook
and possible applications.

\section{The Model}
The model system consists of $N$ individuals(students of a class).
Since the size of a class is not too large, it is suitable to
assume that each student has chances to contact all his
classmates. For clarity, we introduce a generalization of the
$N\times N$ adjacency matrix to describe the interpersonal
relationships of the small social network. The matrix elements
$\omega_{ij}$ represent the weight of edge $e_{ij}$, where $i$,
$j$=1, 2, \ldots, $N$. Postulate that the value of $\omega_{ij}$
is discrete and can be negative for the case of disfavor
relationships. If most elements of the matrix are positive, the
system can be called harmonious; otherwise, it contains
considerable hostility. As an original model, we add an assumption
that each contact can only alter weight by $\pm1$ at most. In
other words, love or hatred is not formed in a day (or individuals
will not fall in ``love" at first sight). This condition makes the
contacts moderate and can be interpreted by the fact that true
friends (or enemies) are selected by time.

Now, take $N$=5 for instance, the adjacency matrix is as below:

$$
\left(%
\begin{array}{ccccc}
0 & \omega_{12} & \omega_{13} & \omega_{14} & \omega_{15} \\
\omega_{21} & 0 & \omega_{23} & \omega_{24} & \omega_{25} \\
\omega_{31} & \omega_{32} & 0 & \omega_{34} & \omega_{35} \\
\omega_{41} & \omega_{42} & \omega_{43} & 0 & \omega_{45} \\
\omega_{51} & \omega_{52} & \omega_{53} & \omega_{54} & 0 \\
\end{array}%
\right)
$$For undirected and weighted graphs, $\omega_{ii}$=0 and
$\omega_{ij}$=$\omega_{ji}$. In the following, we will only
discuss the case of symmetric weights. Since the $i$th row of the
matrix records the information of interpersonal relationships of
student $i$, we will use it to judge the interpersonal similarity.
This point is a basic assumption of our model and will be reviewed
later. At the beginning of evolution, it is reasonable that some
initial weights have non-zero values, due to first impressions
among individuals. For convenience, we assign value 1 with
probability $p$, and $-1$ with probability $1-p$ to the elements
of matrix, i.e. the seed of the model is given. Here, $p$ is
called \emph{the initial amity possibility}. The symmetry
requirement of the matrix must be satisfied. The initial
configuration and the evolution of the system are moderate in
degree. The definition of the model is based on the weights'
dynamics:

(i)First, suppose student $i$ has been randomly selected from the
class. Then, he takes the initial to contact student $j$ with a
certain probability. A natural idea is to set this possibility as:

\begin{equation}
P_{i\rightarrow
j}=\frac{|\omega_{ij}|}{\sum_{j=1}^N|\omega_{ij}|},
\end{equation}recalling the point that ``in the deepest of one's heart are
always those he hates or he loves, while people without
intensified contact are easy to fade from memory"(see
Introduction). However, this method may lead to the absurd case
that individual $j$ with $\omega_{ij}$=0 would not be chosen by
$i$, nor would $i$ by $j$. Therefore, the edges with 0 weight keep
invariant, that is, unfamiliar ones are always unfamiliar. To
avoid this unrealistic scenario, let $i$ choose $j$ with
possibility:

\begin{equation}
W_{i\rightarrow
j}=\frac{|\omega_{ij}|+1}{\sum_{j=1}^N(|\omega_{ij}|+1)},
\end{equation}obviously,
\begin{equation}
W_{i\rightarrow j}=W_{j\rightarrow i}.
\end{equation}

The adaption is based on the moderate evolution mechanism: the
minimum unit of weight is 1, and each contact can only alter
weight by $\pm1$ at most. A non-zero value of $W_{i\rightarrow j}$
is needed but we expect a least deflection to $P_{i\rightarrow
j}$. When $i$ selects $j$ with possibility $W_{i\rightarrow j}$,
for small $|\omega_{ij}|$ this perturbation is quite significant
and reasonable. In a fresh environment, people will try to get
familiar with others and we call it unfamiliar-familiar period.
The differences of weights are not significant; thus, the contacts
between them behave no obvious preferences. Once some weight
becomes 0 during the period, it is still possible to be altered in
later contacts. Initially, the interpersonal relations of the
class are unsteady, and the first impressions rising in the period
will play an important role in future weight polarization. When
friends and enemies (large $|\omega_{ij}|$) form in the system,
situation is quite different and we call it friend-enemy period.
The contacts between friends and encounters between enemies now
become more frequent, and the interpersonal relationships tend to
be steady. However, the emergence of new friends and enemies is
not forbidden.

(ii)Now, $i$ and $j$ have been chosen for interaction, then
$\omega_{ij}$ will be altered with a certain possibility:

\begin{equation}
\omega_{ij}\longrightarrow\omega_{ij}\pm1.
\end{equation}
The crux of the problem now is how to determine the possibility.
Recall that similar human relations and social environments are
more likely to promote friendships. We could define $\gamma_{ij}$
as below to describe the interpersonal relation similarity:

\begin{equation}
\gamma_{ij}=C^{-1}\sum_{\alpha}\omega_{i\alpha}\cdot\omega_{\alpha
j},
\end{equation}where
\begin{equation}
C=\sqrt{\sum_{\alpha}\omega_{i\alpha}^{2}}\cdot\sqrt{\sum_{\beta}\omega_{j\beta}^{2}}.
\end{equation}It is manifest that $\gamma_{ij}$=$\gamma_{ji}$ and
$-1\leq\gamma_{ij}\leq$1. One can see that the definition of
$\gamma_{ij}$ is equivalent to the inner product of two normalized
vectors. So $\gamma_{ij}$ could be regarded as a signed
possibility. Then in detail, the rules are: when
$\gamma_{ij}\geq$0
\begin{eqnarray}
\omega_{ij}\rightarrow\omega_{ij}+1,
\omega_{ji}\rightarrow\omega_{ji}+1
\end{eqnarray}
with possibility $\gamma_{ij}$, and nothing is altered with
possibility $1-\gamma_{ij}$;

when $\gamma_{ij}<$0
\begin{eqnarray}
\omega_{ij}\rightarrow\omega_{ij}-1,
\omega_{ji}\rightarrow\omega_{ji}-1
\end{eqnarray}
with possibility $|\gamma_{ij}|$, and nothing is altered with
possibility $1-|\gamma_{ij}|$.

The mechanism (ii) have simple physical and realistic
interpretations. Take Jack and Mike for instance. If they have
common friends or common enemies, they are more likely to
strengthen their friendship ($\gamma>0$). Suppose, however, they
are good fellows at first and Jack's pals are all Mike's foes. If
Jack goes on associating with Mike, he may be excluded by some of
his friends and have to confront his foes under certain
circumstances. Thus, their social relations have a potential to
separate them ($\gamma<0$), just like an electron-positron system
under external electro-magnetic field. In this case, we can
equally say that Jack and Mike have distinct social tastes and in
the long run, their friendship is on test.

After the weights have been updated, the process is iterated by
randomly selecting a new individual for the next contact, i.e.
going back to step (i) until the class disbands.

To better understand the micro dynamics, it is beneficial to
analyze the case for $N$=3. Suppose Jack(A), Mike(B) and John(C)
interact with each other according to above mechanism. The
possible states of this triangle-relation evolution are shown in
Fig. 1. Triangle (b) represents the friend-friend-enemy relation,
that is, two edges of the triangle have positive(+) weights and
another is negative($-$). However, the common friend of
antagonistic two will play a conciliatory role in the evolution,
and thus, the negative weight will be neutralized at some point,
given sufficient interacting time. Review that the ``enemy" of
Jack's friends or the friend of Jack's ``enemies" may be hard to
associate with Jack (see Introduction). On the basis of similar
analysis, triangle (c) and (d) are expectantly the most possible
stationary states in the evolution. The asymmetry of the micro
mechanism will lead to the asymmetric weight distribution at the
macro level. This point will prove itself in the next section.

\begin{figure}
\scalebox{0.55}[0.55]{\includegraphics{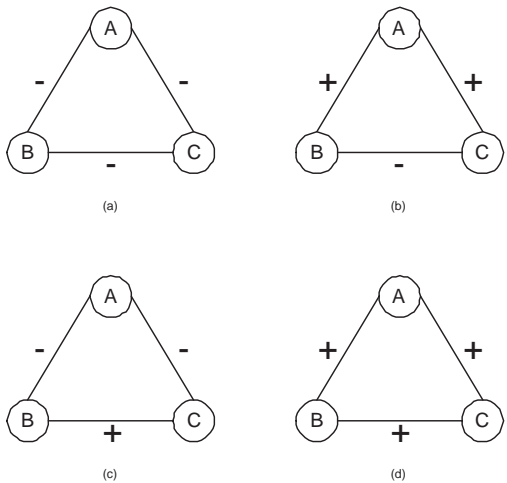}}
\caption{\label{fig:epsart} the possible states of triangle
relationship. Positive(+) edge means friendly relation and
negative($-$) represents hostility.}
\end{figure}

\section{Experimental Results}
We choose different \emph{initial amity possibility} $p$ to
perform simulations. In order to obtain the weight distribution,
the range of weight $\omega$ is equally divided by $M$. Then, the
range [$\omega_{min}$, $\omega_{max}$] becomes [$\omega_{1}$,
$\omega_{2}$), [$\omega_{2}$, $\omega_{3}$), \ldots,
[$\omega_{M}$, $\omega_{M+1}$], where $\omega_{1}$=$\omega_{min}$,
$\omega_{M+1}$=$\omega_{max}$. Define $n_{\omega_{l}}$ as the
number of weights in [$\omega_{l}$, $\omega_{l+1})$, $l$=1, 2,
$\ldots$, $M$; when $l$=$M$ the interval is $[\omega_{M},
\omega_{M+1}]$. Plotted in Fig. 2-7 are typical weight
distributions ($n_{\omega}\sim\omega$) which behave a pinnacle for
$p$=0 or 0.50, power-law with a heavy tail for $p$=0.59 or 0.60,
an exponential decay for $p$=0.70, and a peak structure for
$p$=1.00.

\begin{figure}
\scalebox{0.8}[0.8]{\includegraphics{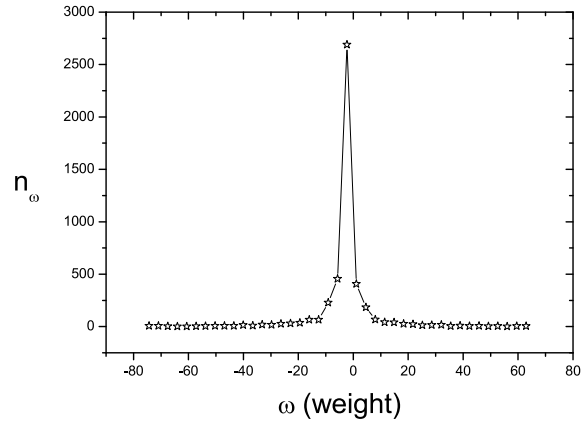}}
\caption{\label{s:p}weight distribution for N=100, p=0 after
$1.0\times10^{6}$ time steps.}
\end{figure}

When $p$=0, i.e. the initial non-diagonal elements are all $-1$,
the final weight distribution exhibits a symmetric pinnacle near
$\omega$=0. The peak value is 2690, and on both sides $n_{\omega}$
is quite low (but many non-zero), see Fig. 2. The weight
distribution for $p\leq$0.50 exhibits a similar behavior, as
experiments can test. For $p$=0.50, the peak value is 1693, see
Fig. 3. When the initial hostility proportion $1-p$ is
significant, the model mechanics has a tendency to push the peak
towards the right. However, no matter how many time steps are run,
the peak cannot move further beyond zero. We can conclude from
these results that when the initial amity is insufficient, the
harmony of the class is out of the question. The most majority are
indifferent to others, and true friends and foes can rarely
``survive" under such environments.

\begin{figure}
\scalebox{0.8}[0.8]{\includegraphics{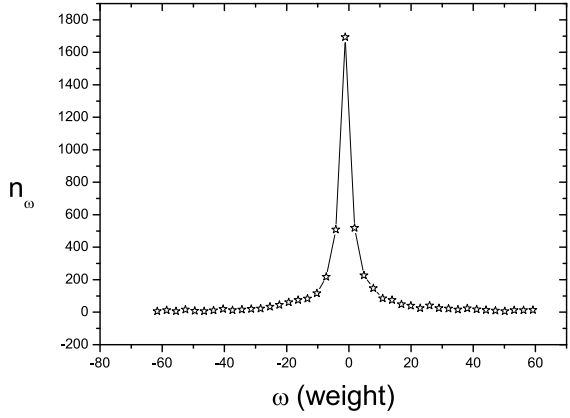}}
\caption{\label{fig:epsart} weight distribution for N=100, p=0.50
after $1.0\times10^{6}$ time steps.}
\end{figure}

\begin{figure}
\scalebox{0.8}[0.8]{\includegraphics{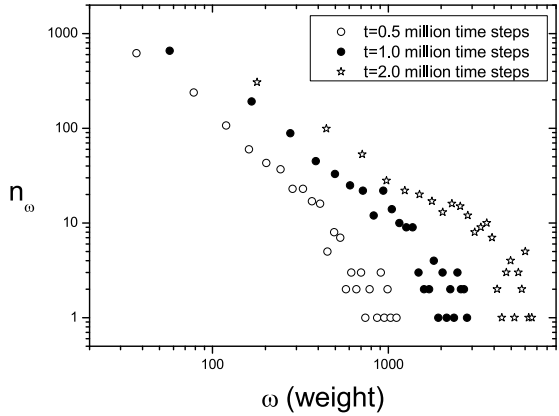}}
\caption{\label{fig:epsart} weight distribution for N=100, p=0.60
after $0.5\times10^{6}$, $1.0\times10^{6}$ and $2.0\times10^{6}$
time steps.}
\end{figure}
Presented in Fig. 4 and Fig. 5 are positive weight distributions
for $p$=0.60 and $p$=0.59. The negative weights are discarded in
the log-log plot. Apparently, each distribution obeys power law
with a heavy tail; this power-law property is independent from
iterated steps. Here, negative weights in the matrix are quite
sparse, compared with the positive, see Fig. 8 and related
discussion.

\begin{figure}
\scalebox{0.8}[0.8]{\includegraphics{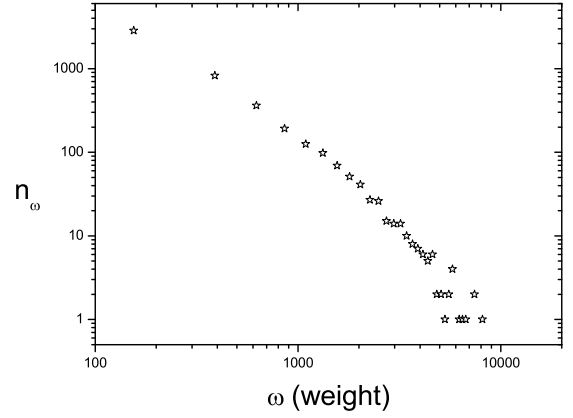} }
\caption{\label{fig:epsart} weight distribution for N=200, p=0.59
after $9.0\times10^{6}$ time steps.}
\end{figure}
Exponential distributions are found near $p$=0.70 and independent
from time steps, as shown in semi-log plot (see Fig. 6). The
negative weights have disappeared under such circumstances. It is
clear that for large $\omega$, $n_{\omega}$ increases with the
passage of time.
\begin{figure}
\scalebox{0.8}[0.8]{\includegraphics{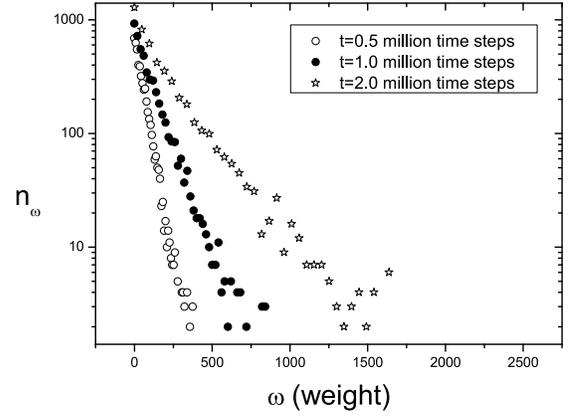}}
\caption{\label{fig:epsart} weight distribution for N=100, p=0.70
after $0.5\times10^{6}$, $1.0\times10^{6}$ and $2.0\times10^{6}$
time steps.}
\end{figure}

\begin{figure}
\scalebox{0.8}[0.8]{\includegraphics{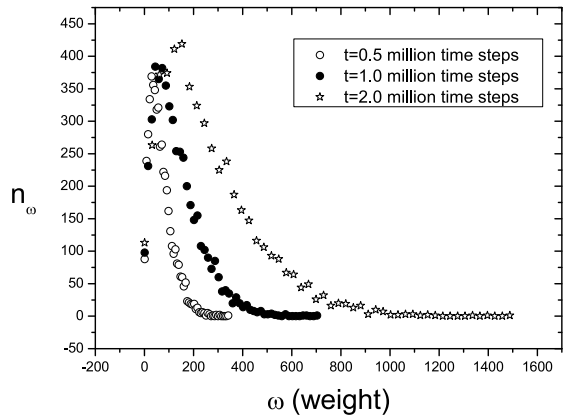}}
\caption{\label{fig:epsart} weight distribution for N=100, p=1.00
after $0.5\times10^{6}$, $1.0\times10^{6}$ and $2.0\times10^{6}$
time steps.}
\end{figure}
When $p$=1.00, i.e. the initial non-diagonal elements are all 1,
we find a peak structure in $n_{\omega}\sim\omega$ diagram(Fig.
7). Different from the above, the maximum of $n_{\omega}$ is
reached somewhat far from $\omega$=0. By increasing the iterated
times, the peak and the upper limit of $\omega$ are both pushed to
the right. It means the harmony of the system is boosted up. One
can check that for $p\geq$0.8, the weight distribution displays
similar behaviors.

Three points must be stressed here. First, we have observed four
typical kinds of distributions from $p$=0 to $p=1.00$, and each
kind appears in a certain range. However, the distributions in
some unmentioned ranges are not so typical and may be influenced
by increasing time steps. Second, by comparing the weight
distributions of different $p$ from 0 to 1, one can see that this
system exhibits a potential to become harmonious. Third, the
properties of the above weight distributions also suggest a
critical transition. Define \emph{the hostility proportion} $h$ as

\begin{equation}
\sum_{i,j,\omega_{ij}<0}\omega_{ij}\big/\sum_{i,j}|\omega_{ij}|
\end{equation}
which can describe the harmony degree of the class from an
opposite sight. The dependence of $h$ on $p$ is shown in Fig. 8,
and a phase transition is found near $p_{c}$=0.6, where the weight
distribution exhibits power-law (Fig. 4 and Fig. 5). Below the
critical value, the hostility proportion $h$ is non-trivial, that
is, there exists considerable hostility in the interpersonal
atmosphere; while above the critical value, the final hostility
proportion $h$ is close to 0, i.e. the interpersonal atmosphere of
the class is quite harmonious. Conflicts and grievances may melt
gradually under such harmonious environment.

\begin{figure}
\scalebox{0.8}[0.8]{\includegraphics{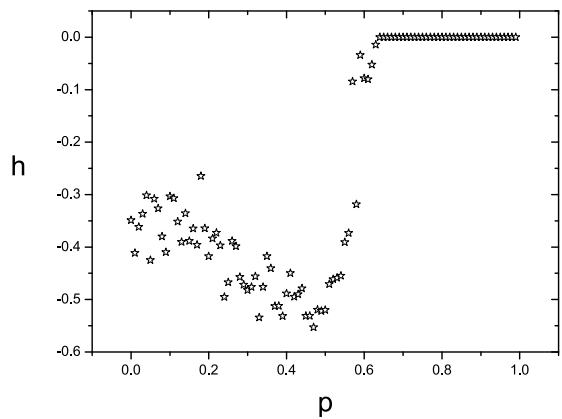}}
\caption{\label{fig:epsart} the dependence of hostility proportion
$h$ on the initial amity proportion $p$, hostility-amity phase
transition for N=100 after $1.0\times10^{6}$ time steps.}
\end{figure}

\section{Review and Outlook}
Of the academic disciplines the social sciences have the longest
history of the substantial quantitative study of real-world
networks \cite{ref9,ref10}. Of particular note among the early
works on the subject are: Jacob Moreno's work in the 1920s and 30s
on friendship patterns within small groups \cite{ref3}; the
so-called "southern women study" of Davis et al. \cite{ref11},
which focused on the social circles of women in an unnamed city in
the American south in 1936; the study by Elton Mayo and colleagues
of social networks of factory workers in the late 1930s in Chicago
\cite{ref12}; the studies of friendship networks of school
children by Rapoport and others \cite{ref4,ref14}; and the
mathematical models of Anatol Rapoport \cite{ref13}, who was one
of the most theorists to stress the significance of the degree
distribution in networks of all kinds, not just social networks.
In more recent years, studies of business communities
 \cite{ref5,ref15,ref16} and of patterns of sexual contacts
 \cite{ref17,ref18,ref19,ref20,ref21} have attracted particular
attention. However, traditional social network studies often
suffer from problems of inaccuracy, subjectivity, and small sample
size. Because of these problems many researchers have turned to
other methods for probing social networks. One source of copious
and relatively reliable data is collaboration networks
\cite{ref22,ref23}; another source of reliable data about personal
connections between people is communication records of certain
kinds \cite{ref24,ref25}. It is quite possible for researchers to
investigate the (weighted) relationships in a certain class or
club, for its finite size and simple patterns. Previous researches
had stressed the significance of the degree distribution in social
networks, while more practical studies on the weighted social
networks are required.

Recently, Alain Barrat, et al. have proposed a general model for
the growth of weighted networks \cite{ref26}, considering the
effect of the coupling between topology and weights' dynamics. It
appears that there is a need for a modelling approach to complex
networks that goes beyond the purely topological point.

The simple model here is self-generated and allows various further
modifications. Some acute interacting ingredients could be taken
into this system. For instance, Jack and Mike might be intimate
friends long before, so the initial weight between them must be
larger. In the present model, moreover, good friendship will not
collapse instantly, nor will old grievance; thus it seems more
reasonable to take some acute interaction into account. In the
framework of this model, it is possible and interesting to study
the adaptive process of a new student joining the class midway
\cite{ref8}. Meanwhile, this model could be easily extended to
directed graph more close to real world where human relations are
often asymmetric. Finally, generalizing it to complex social,
economic and political networks is also an interesting and
challenging task. The relationships between individuals, economic
entities or nations are amazingly similar in many aspects. The
basic assumptions and concepts here are expected to have well
applications in related fields.

\begin{acknowledgements}
This work was supported by the State Key Development Programme of
Basic Research of China (973 Project), the National Natural
Science Foundation of China under Grant No.70271070, the
China-Canada University Industry Partnership Program (CCUIPP-NSFC
No.70142005), and the Doctoral Fund from the Ministry of Education
of China.
\end{acknowledgements}

\end{document}